\documentclass{article}
\usepackage{amsfonts,amsmath,amsxtra,graphicx}
\usepackage{wrapfig}
\usepackage{mathtext}
\usepackage[T2A]{fontenc}
\usepackage{amsmath,amsfonts,amsxtra, mathtools}
\usepackage{amssymb}
\usepackage{amscd}
\usepackage[usenames,dvipsnames]{color}
\usepackage{color}
\usepackage{euscript}
\usepackage[matrix,arrow,curve]{xy}
\usepackage{mathrsfs}
\usepackage{dsfont}
\usepackage{array}
\usepackage{indentfirst}
\usepackage{textcomp}
\usepackage{bbold}

\def\be{\begin{eqnarray}}
\def\ee{\end{eqnarray}}
\def\nn{\nonumber}

\def\0{\emptyset}

\def\Tr{{\rm Tr}}
\def\qTr{{\rm qTr}}
\def\l[{\phantom.[}

\def\lm{\limits}

\def \IC {\mathbb{C}}
\def \IZ {\mathbb{Z}}
\def \IR {\mathbb{R}}

\hoffset= - 1.0in         % switch off for draft style
\begin{document}
\title{{\bf {$SU(2)/SL(2)$ knot invariants and KS monodromies} \vspace{.2cm}}
\author{{\bf D.Galakhov$^{a,b,}$}\footnote{galakhov@itep.ru; galakhov@physics.rutgers.edu}, {\bf A. Mironov$^{a,c,d,e,}$}\footnote{mironov@itep.ru; mironov@lpi.ru} and {\bf A. Morozov$^{a,d,e,}$}\thanks{morozov@itep.ru}}
\date{ }
}

\maketitle

\vspace{-6.0cm}

\begin{center}
\hfill FIAN/TD-09/15\\
\hfill IITP/TH-16/15\\
\hfill ITEP/TH-25/15\\
\end{center}

\vspace{4.2cm}

\begin{center}
$^a$ {\small {\it NHETC and Department of Physics and Astronomy, Rutgers University,
Piscataway, NJ 08855-0849, USA }}\\
$^b$ {\small {\it Institute for Information Transmission Problems, Moscow 127994, Russia}}\\
$^c$ {\small {\it Lebedev Physics Institute, Moscow 119991, Russia}}\\
$^d$ {\small {\it ITEP, Moscow 117218, Russia}}\\
$^e$ {\small {\it National Research Nuclear University MEPhI, Moscow 115409, Russia }}\\
\end{center}

\vspace{1cm}

\begin{abstract}
We review the Reshetikhin-Turaev approach to construction of non-compact knot invariants involving R-
matrices associated with infinite-dimensional representations, primarily those made from Faddeev's quantum
dilogarithm. The corresponding formulas can be obtained from modular transformations of conformal
blocks as their Kontsevich-Soibelman monodromies and are presented in the form of transcendental integrals,
where the main issue is manipulation with integration contours.
We discuss possibilities to extract more
explicit and handy expressions which can be compared with the ordinary (compact) knot polynomials
coming from finite-dimensional representations of simple Lie algebras, with their limits and properties. In
particular, the quantum A-polynomials, difference equations for colored Jones polynomials should be
the same, just in non-compact case equations are homogeneous, while they have a non-trivial right-hand
side for ordinary Jones.
\end{abstract}

\section{Introduction}

At the present stage the most effective way to get
formulas for knot/link invariants \cite{knotpoly,Jones} from CS theory \cite{CS,Wit}
is to use the Reshetikhin-Turaev (RT) formalism \cite{RT,RTmore}.
It arises, e.g., in the temporal gauge \cite{MS} and depends on the oriented link ${\cal L}$
through its projection on $2d$ plane called link diagram ${\cal D}_{\cal L}$
with two types of vertices of valence $(2,2)$.
It provides answers as contractions of quantum ${\cal R}$-matrices
and "turning" matrices which act on a pair of lines and on a single
line in the link diagram respectively.
Independence on the choice of diagram (Reidemeister invariance)
follows from the general properties
of ${\cal R}$ and turning matrices.

Thus, in order to understand which invariants can be obtained via the RT formalisms, one has to enumerate the available numerical quantum ${\cal R}$-matrices and to check if traces of their proper products can be calculated. To this end, first of all, one can use the numerical ${\cal R}$-matrices obtained from the universal quantum ${\cal R}$-matrices of finite-dimensional compact $q$-deformed Lie algebras at finite-dimensional irreducible representations. It immediately gives rise to knot invariants described by finite sums, that is, to knot Laurent polynomials of $q$. Hence, the name compact invariants. Another possibility is to use the ${\cal R}$-matrices for finite-dimensional non-compact $q$-deformed Lie algebras at infinite-dimensional irreducible representations. In this case, one obtains knot invariants represented by integrals. We call these invariants non-compact.
At last, one can consider infinite-dimensional Lie algebras. Though in this case there are good ${\cal R}$-matrices, it is unclear how to define the proper traces of their products. Hence, we restrict ourselves here with the first two possibilities.

Thus, we are going to consider the ${\cal R}$-matrices that can be obtained from
Drinfeld's universal formula, and this is the best choice for
finite-dimensional representations of $SU_q(N)$.
In the case of non-compact $SL_q(N)$ with infinite-dimensional
representations there are at least two interesting
associated quantum ${\cal R}$-matrices:
Faddeev's matrix \cite{Fad} made from quantum dilogarithms and obtained from triangulations of the time slice \cite{GMM},
which was recently used by K.Hikami and R.Inoue to construct knot invariants \cite{Hikn} and another one made from polyhedra, which was used by K.Hikami earlier \cite{Hiko}. They depend on different
numbers of variables, still are intimately related.

These ${\cal R}$-matrices look very similarly and, as we will explain, are obtained from the same quantum algebra $SL_q(N)\otimes SL_{\tilde q}(N)$, where $q=e^{\pi i b^2}$ and $\tilde q=e^{-\pi i b^{-2}}$ with some parameter $b$. Moreover, there are three different ways to proceed: it can be obtained via triangulations ({\bf i}) \cite{Hikn} with the Kontsevich-Soibelman (KS) monodromies \cite{GMM}, and it can be obtained just directly via the Drinfeld double ({\bf ii}), both these ways leading to the same Faddeev ${\cal R}$-matrix. The third way of doing ({\bf iii}) is to add to $SL_q(N)\otimes SL_{\tilde q}(N)$ the second set of Cartan generators to produce the Heisenberg double and then to make a new Drinfeld ${\cal R}$-matrix of it \cite{Kash}. This leads to a direct sum of representations of $SL_q(N)\otimes SL_{\tilde q}(N)$, and in order to produce knot invariants one has to impose a monodromy condition \cite{Hiko} in order to fix the representation.

We consider here the simplest case of the rank 1 groups so that the compact invariants are associated with the quantum algebra $SU_q(2)$, while the non-compact invariants are associated with the product of two quantum algebras $SL_q(2)\otimes SL_{\tilde q}(2)$. It should be related to associating the non-compact invariant with $SL(2,C)$ in \cite{Z}, but this is beyond the scope of this text. Throughout the review, we mainly use as examples the simplest knots: the trefoil  $3_1$ and the figure eight knot $4_1$ (and sometimes other twist knots for an illustration).
The results presented here are basically known, we just collect them in one review.

\bigskip

To fix our notation, throughout the review we use the Pochhammer symbol defined as
\be
(z|q)_{k}=\prod\lm_{n=0}^{k-1}(1-z q^n)\\
(z|q)_{\infty}=\prod\lm_{n=0}^{\infty}(1-z q^n)
\ee
and the quantum dilogarithm defined as
\be
\Phi_b(z)=\frac{(e^{2\pi b (z+i Q/2)}|q^2)_{\infty}}{(e^{2\pi b^{-1} (z-i Q/2)}|\tilde q^2)_{\infty}},\quad q=e^{\pi i b^2},\; \tilde q=e^{-\pi i b^{-2}},\; Q=b+b^{-1},\ \hbar=i\pi b^2
\ee
We also use the $q$-binomial coefficients
\be
\left[\begin{array}{c}
        n\\
        k
    \end{array}\right]_q\equiv{(q^2|q^2)_{n}\over (q^2|q^2)_{k}(q^2|q^2)_{n-k}}={[n]!\over [k]![n-k]!}
\ee
with $q$-numbers defined as
\be
[n]\equiv {q^n-q^{-n}\over q-q^{-1}}
\ee
and the notation
\be
\{x\}\equiv x-{1\over x}
\ee

\section{$SU_q(2)$ and $SL_q(2)$ knot invariants}

As we already explained in the Introduction, one can deal equally well both with compact and non-compact knot invariants. Though our main point of interest here is the non-compact case, we list both of them here and briefly explain what are different ways to produce compact invariants. The remaining sections are devoted to different ways of obtaining non-compact invariants.

\subsection{$SU_q(2)$ invariants and conformal field theory}
\subsubsection{Minimal models}

Using the plat representation of knots and the modular transformation matrices of the minimal models of conformal theory, one can construct the $SU_q(2)$ knot invariants $\mathds{M}{\rm n}(K,r)$ \cite{GMM,GMMM}. To this end, one considers the modular transformations of the four-point spherical conformal block containing the degenerate fields $(1,j+1)$ of the minimal model \cite{CFT}, $B_k[j_1,j_2,j_3,j_4](x)$. It is given by the modular $S$-matrix
\be
B_k[j_1,j_2,j_3,j_4](x)=\sum\limits_{l} S_{kl}\left[\begin{array}{cc}
        j_2 & j_3 \\
        j_1 & j_4
    \end{array}\right]B_l[j_2,j_3,j_4,j_1](1-x)
\ee
and manifestly for one of the degenerate fields being $(1,2)$ it is
    \be
    S\left[\begin{array}{cc}
        1 & j_3 \\
        j_1 & j_4
    \end{array}\right]=\left(
    \begin{array}{cc}
        \frac{\Gamma \left(\frac{{j_1}+1}{b^2}+2\right) \Gamma \left(-\frac{b^2+{j_3}+1}{b^2}\right)}{\Gamma \left(\frac{2 b^2+ {j_1}- {j_3}+ {j_4}+1}{2 b^2}\right) \Gamma \left(-\frac{1- {j_1}+ {j_3}+ {j_4}}{2 b^2}\right)} & \frac{\Gamma \left(\frac{ {j_1}+1}{b^2}+2\right) \Gamma \left(\frac{b^2+ {j_3}+1}{b^2}\right)}{\Gamma \left(\frac{2 b^2+ {j_1}+ {j_3}- {j_4}+1}{2 b^2}\right) \Gamma \left(\frac{4 b^2+ {j_1}+ {j_3}+ {j_4}+3}{2 b^2}\right)} \\
        \frac{\Gamma \left(-\frac{ {j_1}+1}{b^2}\right) \Gamma \left(-\frac{b^2+ {j_3}+1}{b^2}\right)}{\Gamma \left(-\frac{ {j_1}+ {j_3}- {j_4}+1}{2 b^2}\right) \Gamma \left(-\frac{2 b^2+ {j_1}+ {j_3}+ {j_4}+3}{2 b^2}\right)} & \frac{\Gamma \left(-\frac{ {j_1}+1}{b^2}\right) \Gamma \left(\frac{b^2+ {j_3}+1}{b^2}\right)}{\Gamma \left(-\frac{ {j_1}- {j_3}+ {j_4}+1}{2 b^2}\right) \Gamma \left(\frac{2 b^2- {j_1}+ {j_3}+ {j_4}+1}{2 b^2}\right)} \\
    \end{array}
    \right)
    \ee
where the matrix elements are labeled with the indices 0 and 2, and for arbitrary values of all the four fields it is determined from the recursion relation
    \be
    S_{q,q'}\left[\begin{array}{cc}
        r+1 & j_3 \\
        j_1 & j_4
    \end{array}\right]=\sum\lm_{s,p}S_{r+1,s}\left[\begin{array}{cc}
    1 & q \\
    r & j_1
\end{array}\right]
S_{q,p}\left[\begin{array}{cc}
    1 & j_3 \\
    s & j_4
\end{array}\right]
S_{s,q'}\left[\begin{array}{cc}
    r & p \\
    j_1 & j_4
\end{array}\right]
S_{p,r+1}\left[\begin{array}{cc}
    r & 1 \\
    q' & j_3
\end{array}\right]
\ee

Similarly, there is the second modular transformation matrix $T$, which is diagonal \cite{GMM,GMMM} with the diagonal elements
\be
T_{2k}[r,r]=(-1)^{k+1}q^{k(k+1)-(r+1)^2-1}
\ee
Then, one constructs the polynomials of these matrices associated with the corresponding knot invariants:
\be
\mathds{M}(3_1,r)=\sum\lm_{k=0}^r S_{0,2k}\left[\begin{array}{cc}
    r & r \\
    r & r
\end{array}\right] T_{2k}[r,r]^3 S_{2k,0}\left[\begin{array}{cc}
r & r \\
r & r
\end{array}\right]\\
\mathds{M}(4_1,r)=\sum\lm_{k=0}^r \sum\lm_{k'=0}^r S_{0,2k}\left[\begin{array}{cc}
    r & r \\
    r & r
\end{array}\right] T_{2k}[r,r]^2 S_{2k,2k'}\left[\begin{array}{cc}
r & r \\
r & r
\end{array}\right] T_{2k'}[r,r]^{-2} S_{2k',0}\left[\begin{array}{cc}
r & r \\
r & r
\end{array}\right]
\ee

\subsubsection{WZWN theory}

Similarly, one can consider modular transformations in the $\widehat {SU(2)}_k$ WZWN theory and find the corresponding $S$- and $T$-matrices \cite{AG} in order to construct the WZWN polynomials \cite{inds}: $\mathds{I}{\rm nd}(K,\rho)$. The matrix $S$ is given just by the quantum $SU_q(2)$ Racah coefficients (6j-symbols) \cite{Racah,KR} and in the case of our interest, is equal to
\be
    \Sigma_{2k,2k'}\left[\begin{array}{cc}
        s & s \\
        s & s
    \end{array}\right]=\frac{\int\lm_1^{\infty}d_q x\; x^{k-k'-s-2}\;{}_2\phi_1\left[\begin{array}{c}
        -k\;-k\\
        -2k\\
    \end{array}\right](x^{-1}){}_2\phi_1\left[\begin{array}{c}
    -k'\;-k'\\
    -2k'\\
\end{array}\right](x)}{\int\lm_1^{\infty}d_q x\; x^{-2(k'+1)}\;{}_2\phi_1\left[\begin{array}{c}
-k'\;-k'\\
-2k'\\
\end{array}\right](x)^2}
    \ee
    Note that
    \be
    \boxed{\Sigma\neq S}
    \ee
The very polynomials are defined in the same way:
    \be
    \mathds{I}(3_1,r)=\sum\lm_{k=0}^r \Sigma_{0,2k}\left[\begin{array}{cc}
        r & r \\
        r & r
    \end{array}\right] T_{2k}[r,r]^3 \Sigma_{2k,0}\left[\begin{array}{cc}
    r & r \\
    r & r
\end{array}\right]\\
\mathds{I}(4_1,r)=\sum\lm_{k=0}^r \sum\lm_{k'=0}^r \Sigma_{0,2k}\left[\begin{array}{cc}
    r & r \\
    r & r
\end{array}\right] T_{2k}[r,r]^2 \Sigma_{2k,2k'}\left[\begin{array}{cc}
r & r \\
r & r
\end{array}\right] T_{2k'}[r,r]^{-2} \Sigma_{2k',0}\left[\begin{array}{cc}
r & r \\
r & r
\end{array}\right]
\ee

\subsubsection{Jones polynomials and closed braids}

Canonically, the knot invariants related to group $SU(2)$ are Jones polynomials \cite{Jones,Wit}, which can be defined in the simplest way by the skein relations \cite{skein} in the fundamental representation and by further cabling \cite{cab} in higher spin representations. Another possibility is to use the RT procedure for the closed braid, with the braid group given by the $SU_q(2)$ R-matrix \cite{RT,RTmore} (for a combination of conformal block and RT calculations see \cite{RTmod}).

The colored reduced Jones polynomials $\mathds{J}(K,r)$ are
    \be\label{Jtr}
    \mathds{J}(3_1,r)=\sum\lm_{j=0}^{\infty} (-1)^j q^{j(j+3)}\prod\lm_{i=1}^j (q^{r+1-i}-q^{-(r+1-i)})(q^{r+1+i}-q^{-(r+1+i)})\\\label{Jfe}
    \mathds{J}(4_1,r)=\sum\lm_{j=0}^{\infty} \prod\lm_{i=1}^j (q^{r+1-i}-q^{-(r+1-i)})(q^{r+1+i}-q^{-(r+1+i)})\\
    \mathds{J}(5_2,r)=\sum\lm_{n=0}^{r-1}\frac{\prod\lm_{j=-n}^n\{r-1+j\}}{ \{r-1\}}(-1)^n\sum\lm_{k=0}^n q^{k^2-k (3 n+2)+n (3 n+5)}\left[\begin{array}{c}
        n\\
        k
    \end{array}\right]_q
    \ee
In general for $k$-th twist knot the Jones polynomial is \cite{Gar,evo}
\be
\mathds{J}_{r-1}^{(k)} ={1\over [r+1]} \sum_{s=0}^{r-1}  F^{(k)}_s\prod_{j=-s}^{s} \{q^{r+j}\}
\ee
with
\be
F^{(k)}_s = q^{s(s+3)/2} \sum_{j=0}^s \frac{\{q^{2j+1}\}\{q^{j+1}\}^{2jk}}
{\prod_{i=j-1}^{j-1+s} \{q^{i+2}\}}
\ee

\subsubsection{Jones polynomials and conformal block calculations}

Naively, these three types of polynomials do not obligatory have to coincide. For instance, the matrices $S$ in the minimal model and WZWN approaches are different:
\be
\boxed{\Sigma=USU^{\dag}}
\ee
where $U$ is diagonal \cite{GMM}. However, it turns out that they coincide, and the diagonal matrix $U$ omits from all answers, i.e.
$\mathds{M}(K,r)=\mathds{I}(K,r)$.

Similarly, one can check their identity with the Jones polynomials. For instance, since
\be
\Sigma_{2k,0}\left[\begin{array}{cc}
    r & r \\
    r & r
\end{array}\right]=(-1)^{k-1}\frac{[k]!^2[s+1+k]![s-k]!}{[2k]![s+1]!^2}
\nn\\
\Sigma_{0,2k}\left[\begin{array}{cc}
    r & r \\
    r & r
\end{array}\right]=(-1)^{k-1}\frac{[2k+1]![s]!^2}{[k]!^2[s-k]![s+1+k]!}
\ee
\be
\mathds{I}(3_1,r)=\frac{q^{3s(s+1)}}{[s+1]^2}\sum\lm_{k=0}^s [2k+1](-1)^kq^{-3k(k-1)}
\ee
and one can verify that
\be
\boxed{\mathds{I}(3_1,r)=\mathds{J}(3_1,r)}
\ee
and, similarly, that
\be
\boxed{\mathds{I}(4_1,r)=\mathds{J}(4_1,r)}
\ee

\subsection{$SL_q(2)$ invariants\label{nc}}

$SL_q(2)$ invariants obtained in different ways ({\bf i}), ({\bf ii}) or ({\bf iii}) turn out to be the same, we denote them $\mathds{H}(K,\rho)$. These are the
integral state models (see, e.g., \cite{Hiko,Hikn,Z,AK}):
    \be\label{Htr}
    \mathds{H}(3_1,x)=e^{{3x^2\over \hbar}+{i\pi x\over \hbar}}
    \ee
    \be\label{Hfe}
      \mathds{H}(4_1,x)=\int dy \frac{\Phi_b(x-y)}{\Phi_b(y)}e^{2\pi i x(2y-x)}
      \ee
      \be
      \mathds{H}(5_2,x)=\int dz \frac{e^{\pi i (z^2-x^2)}}{\Phi_b(z)\Phi_b(z-x)\Phi_b(x+z)}
    \ee

\section{R-matrices for $SL_q(2)$}

Now we are going to explain how the non-compact knot invariants of ss.\ref{nc} can be obtained within the RT formalism. The first ingredient of the approach is the ${\cal R}$-matrix. In this section we briefly describe three ways of obtaining the corresponding ${\cal R}$-matrices mentioned in the Introduction.

\subsection{R-matrix from KS monodromies: summary from \cite{GMM}}

The first method ({\bf i}) to obtain the non-compact ${\cal R}$-matrix was discussed in detail in \cite{GMM}, where we explained how one can construct the ${\cal R}$-matrix for the $n$-strand braid via KS monodromies. The procedure basically consists of a few steps.

\begin{enumerate}
\item First of all, one has to construct the spectral curve:
\be
\lambda^2=\sum\lm_{i=1}^{n}\left(\frac{c_2(R_i)}{(z-x_i)^2}+\frac{u_i}{z-x_i}\right)
\ee

There are {\bf relations} between moduli $u_i$ so there are $n-3$ free moduli for the $n$-strand braid. But we {\bf increase} the number of moduli. This should not affect the problem, we are going to integrate over them eventually, the only information we use is that the residues are fixed at the initial stage:
\be
\oint\lm_{x_i}\lambda=\pm \sqrt{c_2(R_i)}
\ee
Afterwards neither do we keep track of this information as the braid evolves.
So, practically, we start with the following curve:
\be
\lambda^2=\frac{\prod\lm_{j=1}^{2n+1}(z-p_j)}{\prod\lm_{i=1}^{n}(z-x_i)^2}
\ee
Ultimately, we increase the number of moduli $p_i$ by one so the curve reads (to unglue top and bottom tips of triangles):
\be
\boxed{\lambda^2=\frac{\prod\lm_{j=1}^{2n+2}(z-p_j)}{\prod\lm_{i=1}^{n}(z-x_i)^2}}
\ee
If one puts all the $p_i$ on the real axis and trace out a permutation of two singularities, one gets the following picture of WKB lines' transformation depicted at fig.\ref{fig1} (the blue dots mark zeroes of discriminant, while the red dots mark singularities).

\begin{figure}[htbp]
    \begin{center}
        \includegraphics[scale=0.5]{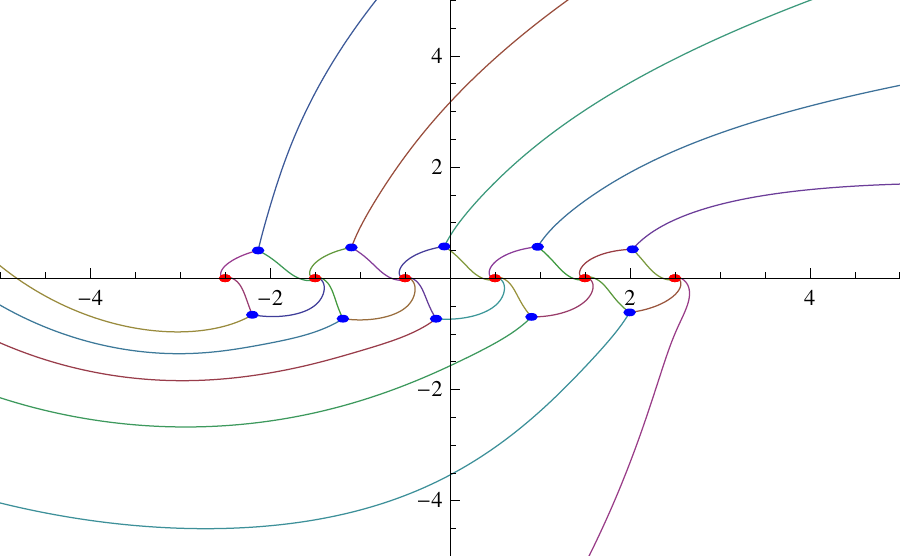}\; \; \;      \includegraphics[scale=0.5]{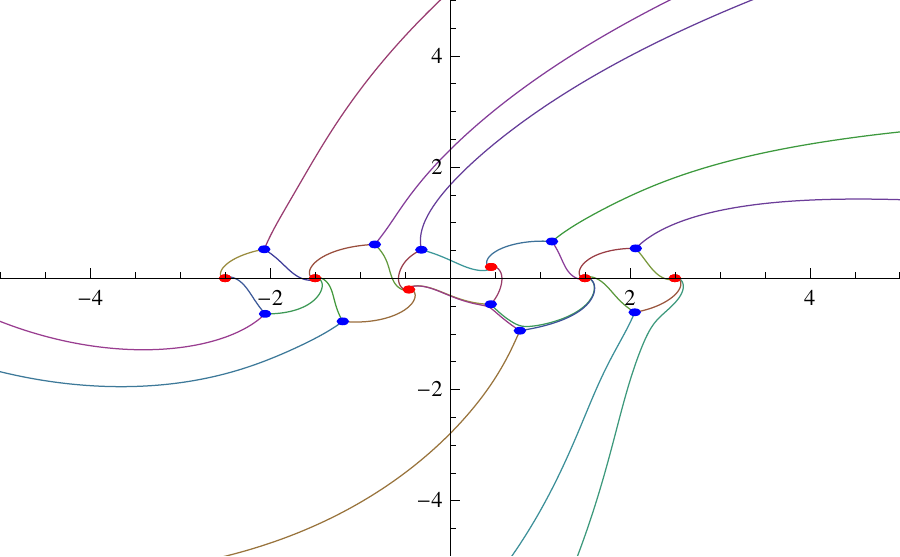}

        \bigskip

        \includegraphics[scale=0.5]{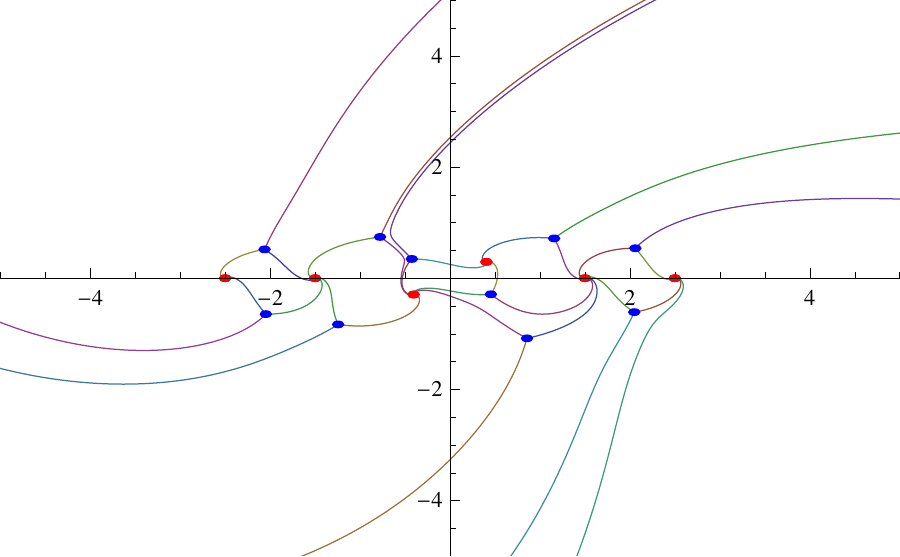}\; \; \;      \includegraphics[scale=0.5]{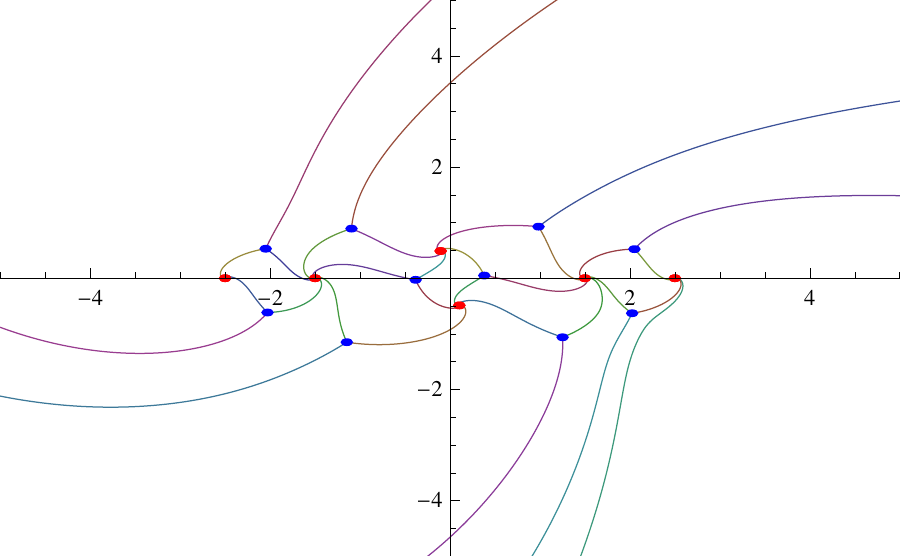}

        \bigskip

        \includegraphics[scale=0.5]{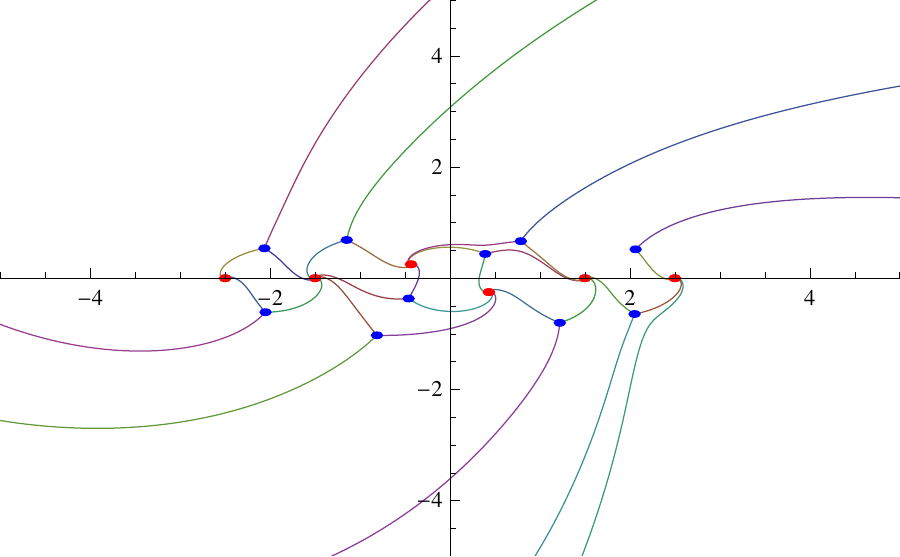}\; \; \;      \includegraphics[scale=0.5]{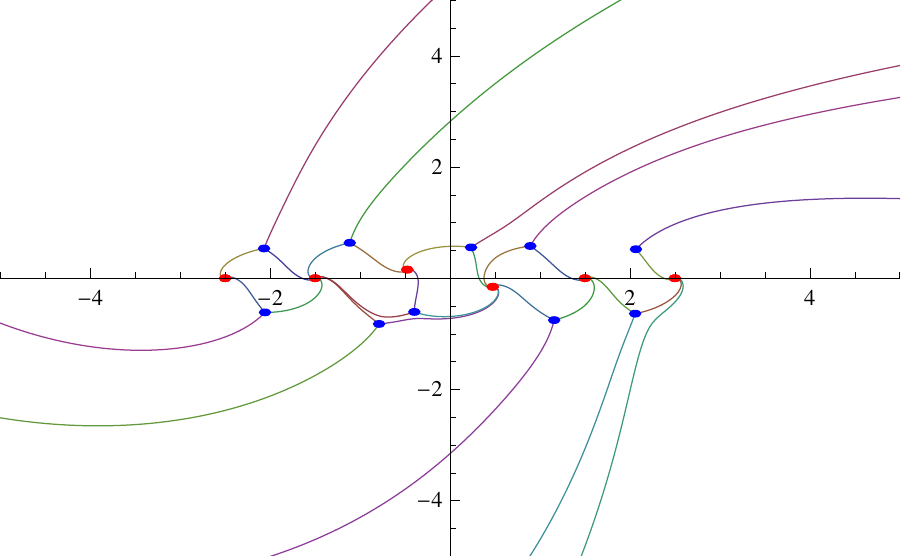}

        \bigskip

        \includegraphics[scale=0.5]{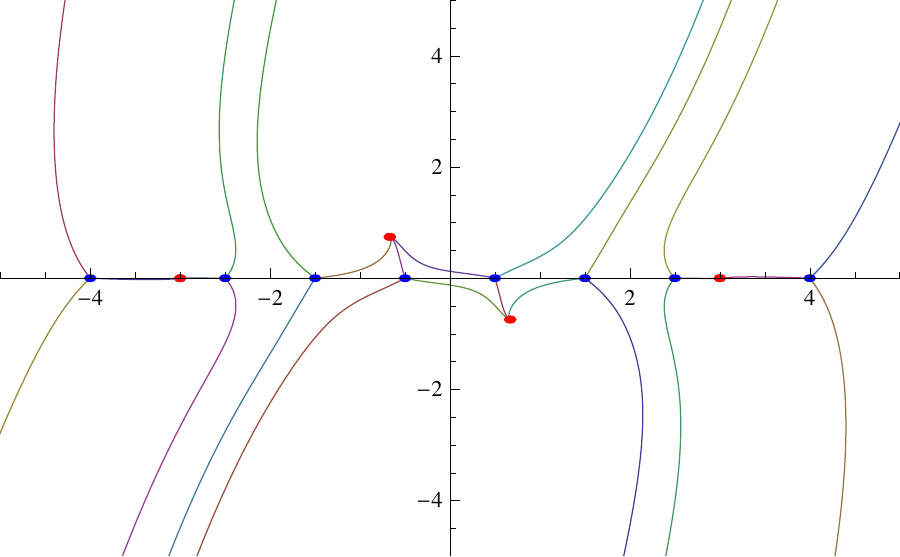}\; \; \;      \includegraphics[scale=0.5]{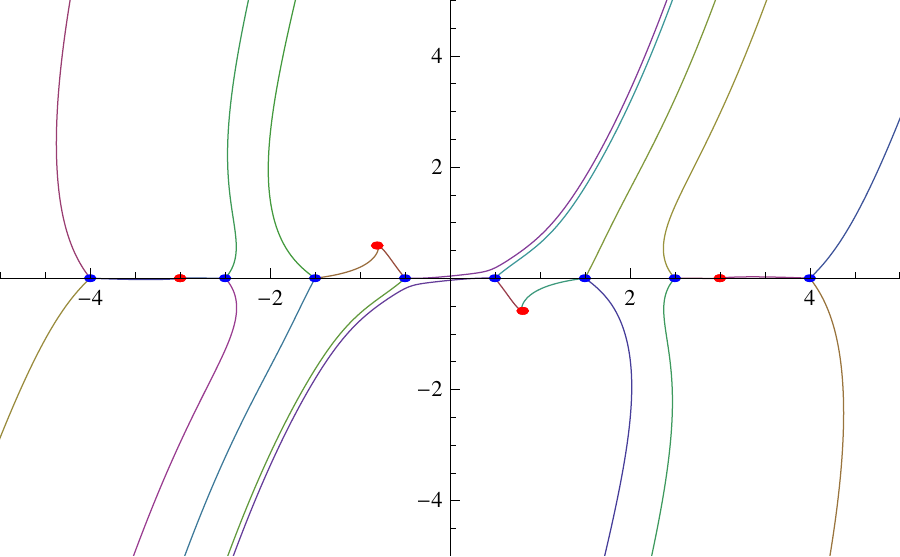}

        \bigskip

        \includegraphics[scale=0.5]{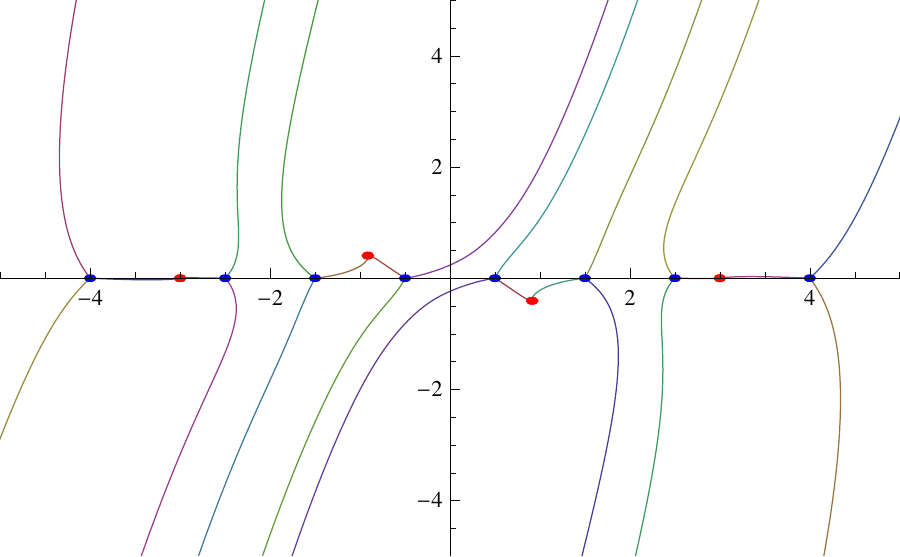}\; \; \;      \includegraphics[scale=0.5]{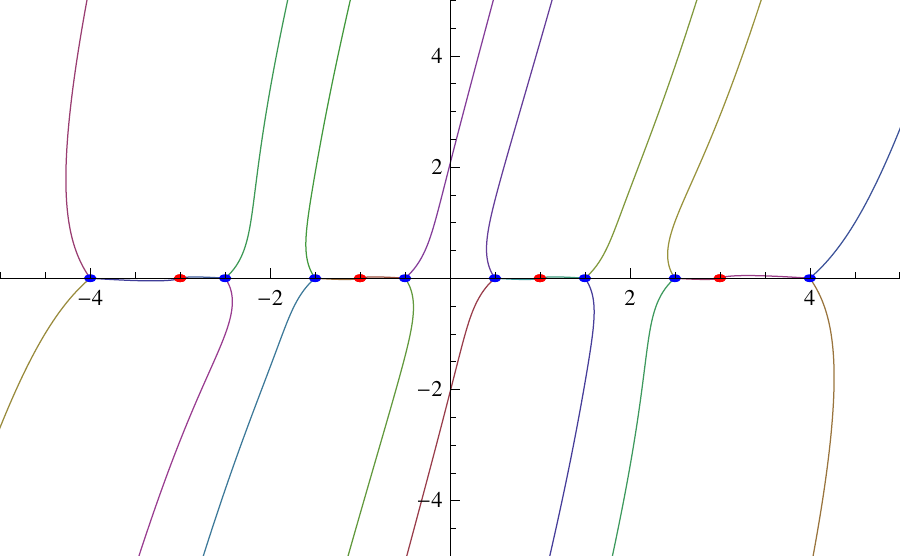}
        \caption{Evolution of the WKB lines. \label{fig1}}
    \end{center}
\end{figure}

The corresponding transformation of triangulations are depicted at fig. \ref{fig2} for two strands. The green edges mark the initial edges of triangulations, while the red edges mark the mutated ones.

\begin{figure}[htbp]
    \begin{center}
        \includegraphics[scale=1]{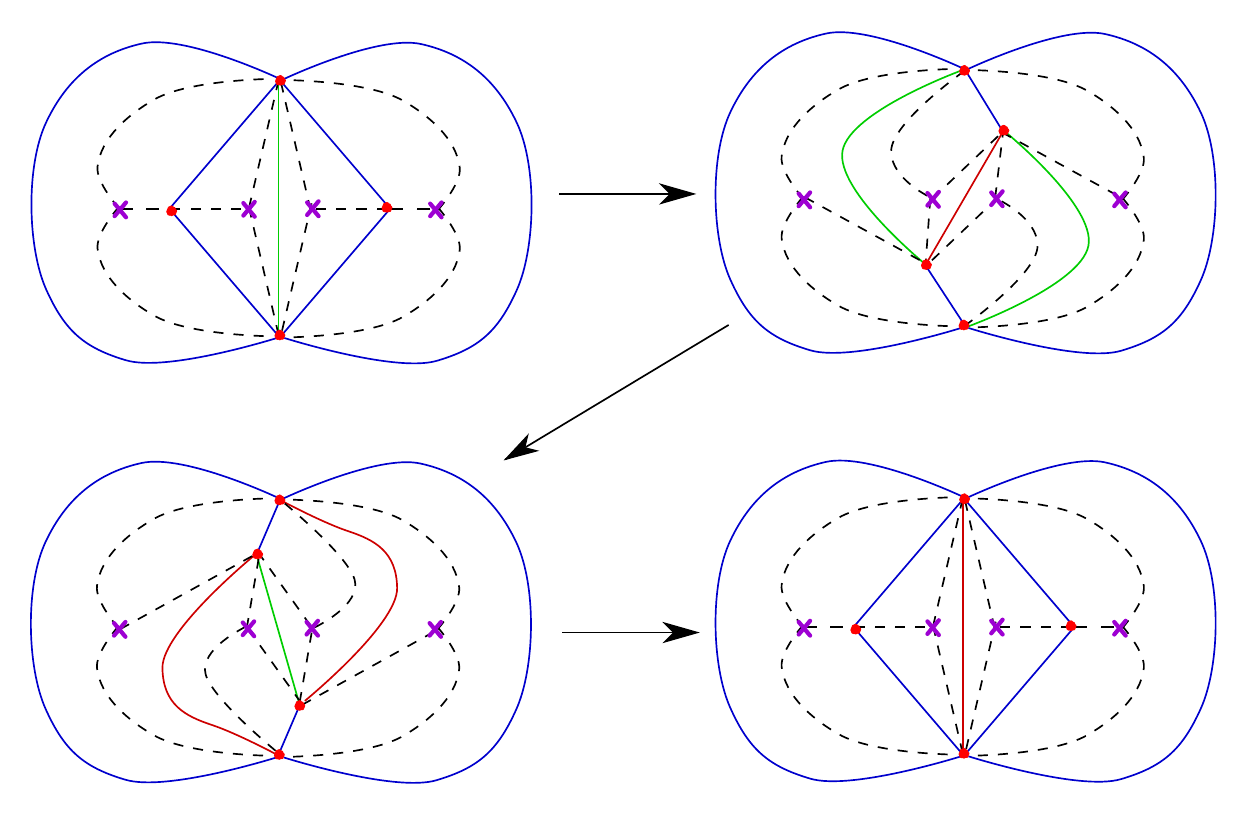}
        \caption{Evolution of the triangulation. \label{fig2}}
    \end{center}
\end{figure}

    \item One considers the "extended" moduli space ${\cal M}$, with coordinates $(x,u,b)$.
    \item One introduces a triangulation flip that corresponds to intersection of a wall, i.e. $\exists\gamma\in H_1(\Sigma),\; \mathop{\rm Im} b^{-1}\lambda|_{\gamma}=0 $. This flip can be manifestly constructed and can be
    \begin{enumerate}
        \item in the Schr\"odinger representation (fundamental action): $\hat\kappa_{\gamma}\sim (\hat w_{\gamma}|q)_{\infty}$
        \item in the Heisenberg representation (adjoint action): $\hat K_{\gamma}O=\hat\kappa_{\gamma}^{-1}O \hat\kappa_{\gamma}$
    \end{enumerate}
    \item The KS invariants are constructed from the flatness condition
    \be
    \prod\lm_{\substack{\rm contractible\\ \rm loop}}^{\leftarrow}\hat K_{\gamma}=\hat {\mathbb 1}
    \ee
    Cut this loop, then, for example, ${\mathds S}=\hat K_{\gamma_2}\hat K_{\gamma_1}=\hat K_{\gamma_1}\hat K_{\gamma_1+\gamma_2}\hat K_{\gamma_2}$, thus ${\mathds S}$ is an invariant across the marginal wall.

    \item Within this framework, one can describe knots as KS monodromies. In fact, the knot can be described as some monodromy of conformal blocks \cite{Wit} in such a way that one considers an initial conformal block whose points evolve permuting with a non-trivial monodromy so that ultimately one has the same conformal block (with possibly permuted points). A point in $\cal M$ is defined by the conformal block. Thus, the knot can be associated to some elements of $\pi_1({\cal M})$.
    \item One constructs an elementary building block for a simple flip
    \be\label{RKS}
    R\sim(\hat w_{\gamma_1}|q^2)_{\infty}(\hat w_{\gamma_2}|q^2)_{\infty}(\hat w_{\gamma_3}|q^2)_{\infty}(\hat w_{\gamma_4}|q^2)_{\infty}\in U(Heis)
    \ee
    \item One constructs the whole ${\cal R}$-matrix element by simple gluing
    \be
    {\cal R}=\prod\lm_i R_i\in \pi_1({\cal M},U(Heis))
    \ee

\end{enumerate}

The ${\cal R}$-matrix (\ref{RKS}) coincides with the ${\cal R}$-matrix obtained by L.Faddeev \cite{Fad}, R.Kashaev \cite{Kash2} and later by K.Hikami \cite{Hikn}. It depends on two extra free constants $c'$, $c''$ (notice that in Hikami's paper the quantum dilogarithm is defined as inverted one):
    \be\label{HikR}
    \begin{split}
    \langle x_1,x_2| R^{(H)}|y_1,y_2\rangle=\frac{\Phi_b\left(x_1-y_1+\frac{i Q}{2}\right)\Phi_b\left(y_2-x_2+\frac{i Q}{2}\right)}{\Phi_b\left(x_1-x_2\right)\Phi_b\left(y_2-y_1\right)}\times \\
    \times e^{2\pi i \left(\frac{iQ}{2}(y_1-y_2-x_1+x_2)+c'(y_2-x_1)+c''(-y_1+x_2)+\frac{1}{12}(1+Q^2)-\frac{1}{2}(c'+c'')^2\right)}
    \end{split}
    \ee

\subsection{R-matrix from PT representation}

Within the second approach ({\bf ii}), one constructs the same ${\cal R}$-matrix for the quantum algebra $SL_q(2)\otimes SL_{\tilde q}(2)$ \cite{BT}.
The universal expression for the ${\cal R}$-matrix reads
\be
{\cal R}=q^{H\otimes H}g_b\left(4(\sin \pi b^2)^2 E\otimes F\right) q^{H\otimes H}
\ee
where
\be
g_b(x):=-\int\lm_{\IR+i 0}\frac{dt}{t}\frac{e^{tQ/2}x^{\frac{t}{2\pi i b}}}{(1-e^{bt})(1-e^{t/b})}
\ee
The representations of $SL_q(2)\otimes SL_{\tilde q}(2)$ are enumerated by two spins $(m,n)$ or one complex
Liouville momentum $\alpha_{m,n}=i(m+1)b + i(n+1)/b$. If one considers a singlet in the second factor $SL_{\tilde q}(2)$, one remains just with the usual Drinfeld $SL_q(2)$ ${\cal R}$-matrix
\be
{\cal R}=q^{H\otimes H}\sum\lm_{n=0}^{\infty}\frac{q^{\frac{1}{2}(n^2-n)}}{[n]_q!}\left((q-q^{-1})E\otimes F\right)^n q^{H\otimes H}
\ee
One can introduce a representation for a continuous  "spin" $s$ on the space of functions of $x$, $L^2(\mathds{R})$:
\be\label{reps}
\rho_s(E)=e^{\pi b x}\frac{\cosh \pi b(p-s)}{\sin \pi b^2}e^{\pi b x},\quad  \rho_s(F)=e^{-\pi b x}\frac{\cosh \pi b(p+s)}{\sin \pi b^2}e^{-\pi b x},\quad \rho_s(q^H)=e^{-\pi b p}
\ee
where $[p,x]=(2\pi i)^{-1}$. Notice that the trivial (scalar) representation appears at $s=-i Q/2$. Afterwards we can project the ${\cal R}$-matrix on the representation $s_2 \otimes s_1$ and construct its action on a representative $\psi(x_2,x_1)$. This allows one to represent the ${\cal R}$-matrix in the form of an integral kernel:
\be
\begin{split}
R_{s_2,s_1}=(\rho_{s_2}\otimes \rho_{s_1}){\cal R}\\
(R_{s_2,s_1}\psi)(x_2,x_1)=\int dy_1 dy_2\; R_{s_2,s_1}\left[\begin{array}{cc}
    y_2 & y_1\\
    x_2 & x_1 \\
\end{array}\right] \psi(y_2,y_1)
\end{split}
\ee
An explicit expression for this integral kernel is given in \cite{BT}
\be\label{R-op-Tesch}
\begin{split}
R_{s_2,s_1}\left[\begin{array}{cc}
    y_2 & y_1\\
    x_2 & x_1 \\
\end{array}\right] = e^{2 i \pi  \left(\frac{Q^2}{4}+\frac{1}{2} i Q \left(-x_1+x_2-y_1+y_2\right)+s_1 \left(y_1-x_1\right)+s_2 \left(x_2-y_2\right)+s_1 s_2\right)}\times\\ \times\frac{G_b\left(\frac{Q}{2}+\frac{i}{2}  \left(s_1+s_2\right)+i \left(x_2-x_1\right)\right) G_b\left(\frac{Q}{2}-\frac{i}{2}  \left(s_1+s_2\right)+i \left(y_2-y_1\right)\right)}{G_b\left(Q+\frac{i}{2}  \left(s_1-s_2\right)+i \left(x_2-y_1\right)\right) G_b\left(Q-\frac{i}{2}  \left(s_1-s_2\right)+i \left(y_2-x_1\right)\right)}
\end{split}
\ee
where
\be
G_b(x)\sim\frac{\prod\lm_{n=1}^{\infty}(1-e^{2\pi i b^{-1}(x-n b^{-1})})}{\prod\lm_{n=0}^{\infty}(1-e^{2\pi i b (x+n b)})}= \frac{1}{\Phi_b\left(i\left(x-\frac{Q}{2}\right)\right)}
\ee
Comparing expression (\ref{R-op-Tesch}) with expression (\ref{HikR}), one concludes that
\be
\boxed{R_{s_2,s_1}\left[\begin{array}{cc}
        y_1+\frac{s_1}{2} & y_2-\frac{s_2}{2}\\
        x_2-\frac{s_1}{2} & x_1+\frac{s_2}{2} \\
    \end{array}\right]=e^{\frac{1}{6} i \pi  \left(2 Q^2+6 s_1^2+6 s_2^2-1\right)}\langle x_1,x_2| R^{(H)}|y_1,y_2\rangle\Bigg|_{\substack{c'=s_1\\ c''=s_2}}}
\ee

The expression for the ${\cal R}$-matrix in the momentum space reads \cite{BT}
\be\label{HF}
R_{s_2,s_1}\left[\begin{array}{cc}
    p_2 & p_1\\
    k_2 & k_1 \\
\end{array}\right]=\delta(p_2+p_1-k_2-k_1)\frac{e^{-i\pi (p_1 k_2+p_2 k_1)}}{G_b(Q+i(p_1-k_1))}\frac{w_b(s_1+k_1)}{w_b(s_1+p_1)}\frac{w_b(s_2-k_2)}{w_b(s_2-p_2)}
\ee
Analogously
\be
R_{s_2,s_1}^{-1}\left[\begin{array}{cc}
    p_2 & p_1\\
    k_2 & k_1 \\
\end{array}\right]=\delta(p_2+p_1-k_2-k_1)\frac{e^{-\pi Q(p_1-k_1)+i\pi (p_1 p_2+k_1 k_2)}}{G_b(Q+i(p_1-k_1))}\frac{w_b(s_1+k_1)}{w_b(s_1+p_1)}\frac{w_b(s_2-k_2)}{w_b(s_2-p_2)}
\ee
where
\be
w_b(x)=e^{\frac{\pi i}{2}(\frac{Q^2}{4}+x^2)}G_b\left(\frac{Q}{2}-ix \right)
\ee
So the relation between the ${\cal R}$-matrix and inverse one reads
\be
\int dk_1' dk_2'\; R_{s_1,s_2}\left[\begin{array}{cc}
    k_1' & k_2'\\
    k_1 & k_2 \\
\end{array}\right]R_{s_1,s_2}^{-1}\left[\begin{array}{cc}
k_1'' & k_2''\\
k_1' & k_2' \\
\end{array}\right]=\delta(k_1-k_1'')\delta(k_2-k_2'')
\ee

\subsection{R-matrix and modular double}

At last, the third way ({\bf iii}) to obtain the ${\cal R}$-matrix in order to generate non-compact knot invariants is due to R.Kashaev \cite{Kash}. It consists of adding another Cartan generator $\bar H$ making the Heisenberg double, and then making the standard Drinfeld ${\cal R}$-matrix element out of it. The general construction looks as follows.

If there exists a Hopf algebra $\cal A$ generated by a basis $\{e_{\alpha}\}$, there is a multiplication rule given by structure constants $m_{\alpha\beta}^{\gamma}$:
\be
e_{\alpha}e_{\beta}=m^{\gamma}_{\alpha\beta}e_{\gamma}
\ee
and a co-product:
\be
\Delta(e_{\alpha})=\mu_{\alpha}^{\beta\gamma} e_{\beta}\otimes e_{\gamma}
\ee
One can consider also the dual algebra ${\cal A}^*$ with the corresponding structures. This algebra spans the dual generators $\{e^{\alpha}\}$. These bases can be joined into one algebra called Heisenberg double $H({\cal A})$ determined by the following relations:
\be
e_{\alpha}e_{\beta}=m^{\gamma}_{\alpha\beta}e_{\gamma},\quad  e^{\alpha}e^{\beta}=\mu_{\gamma}^{\alpha\beta}e^{\gamma},\quad e_{\alpha} e^{\beta}=m_{\rho\gamma}^{\beta}\mu_{\alpha}^{\gamma\sigma}e^{\rho}e_{\sigma}
\ee
There is a canonical element
\be
S=e_{\alpha}\otimes e^{\alpha}
\ee
that satisfies the {\bf pentagon} relation
\be
S_{12}S_{13}S_{23}=S_{23}S_{12}
\ee
Note that in this case the co-product can not be extended to the whole Heisenberg algebra. In order to obtain a Drinfeld double, however, one can embed it into the product of two Heisenberg doubles. Indeed, let us multiply $H(A)$ by the Heisenberg double of a dual algebra $H({\cal A}^*)$ defined as
\be
\tilde e_{\alpha}\tilde e_{\beta}=m^{\gamma}_{\alpha\beta}\tilde e_{\gamma},\quad  \tilde e^{\alpha}\tilde e^{\beta}=\mu_{\gamma}^{\alpha\beta}\tilde e^{\gamma},\quad \tilde e^{\beta} \tilde e_{\alpha}=\mu_{\alpha}^{\sigma\gamma}m_{\gamma\rho}^{\beta}\tilde e_{\sigma} \tilde e^{\rho}
\ee
The canonical element in this algebra is
\be
\tilde S=\tilde e_{\alpha}\otimes \tilde e^{\alpha}
\ee
Then one can construct an associative algebra called Drinfeld double $D({\cal A})$ generated by the elements $\{E_{\alpha},E^{\beta}\}$:
\be
E_{\alpha}E_{\beta}=m^{\gamma}_{\alpha\beta}E_{\gamma},\quad  E^{\alpha}E^{\beta}=\mu_{\gamma}^{\alpha\beta}E^{\gamma}, \quad \mu^{\sigma\gamma}_{\alpha}m^{\beta}_{\gamma\rho}E_{\sigma} E^{\rho}= m^{\beta}_{\rho\gamma}\mu^{\gamma\sigma}_{\alpha}E_{\sigma} E^{\rho}
\ee
The canonical element of the Drinfeld double
\be
R=E_{\alpha}\otimes E^{\alpha}
\ee
satisfies the {\bf Yang-Baxter} equation
\be
R_{12}R_{13}R_{23}=R_{23} R_{13} R_{12}
\ee
One can construct a map:
\be
\phi: \quad D({\cal A})\to H({\cal A})\otimes H({\cal A}^*)
\ee
or explicitly
\be
\phi:\quad E_{\alpha}\mapsto \mu_{\alpha}^{\beta\gamma}\; e_{\beta}\otimes \tilde e_{\gamma},\quad E^{\alpha}\mapsto m^{\alpha}_{\gamma\beta}\; e^{\beta}\otimes \tilde e^{\gamma}
\ee
This map gives the desired quartic factorization formula
\be\label{Hiko}
R^{\phi}_{12,34}=(\phi\otimes \phi) R=\left(S_{14}^{t_4}\right)^{-1} S_{13}\; S_{24}^{t_2t_4} \left(S_{23}^{t_2}\right)^{-1}
\ee

Now consider the example of $SL_q(2)$.
In this case of $SL_q(2)$ as an algebra $\cal A$, one can choose the Borel subalgebra, then the Heisenberg double $H({\cal A})$ is given by the generators $H$, $\bar H$, $E$ and $F$ subject to the following relations ($q=e^{-h}$, $K=q^H$)
\be
\begin{array}{lll}
\left[H,\bar H\right]=1, & \left[E,\bar H\right]=0, & \left[H,E\right]=E\\
\left[H,F\right]=-F & \left[\bar H,F\right]=\hbar F, & \left[E,F\right]=(1-q)K^{-1} \\
\end{array}
\ee
The corresponding basis vectors read
\be
e_{m,n}=\frac{H^m E^n}{m! \prod\lm_{j=1}^{n}(1-q^j)},\quad e^{m,n}=\bar H^m F^n,\quad m,n\in \IZ_{\geq 0}
\ee
Thus, the canonical element reads\footnote{
	Note that
	$$
	(z|q)_{\infty}=\sum\lm_{n=0}^{\infty}\frac{(-1)^n q^{\frac{n(n-1)}{2}}}{\prod\lm_{j=0}^n(1-q^j)}z^n,\quad 	(z|q)_{\infty}^{-1}=\sum\lm_{n=0}^{\infty}\frac{1}{\prod\lm_{j=0}^n(1-q^j)}z^n
	$$
	}
\be
S=\sum\lm_{m,n=0}^{\infty} e_{m,n}\otimes e^{m,n}=e^{H\otimes \bar H}(E\otimes F|q)_{\infty}^{-1}
\ee
One can choose a representation for our algebra:
\be
\begin{split}
H=\frac{\hat q}{h},\quad \bar H=\hat p,\quad E=q^{\frac{1}{4}}e^{\hat p},\quad F=q^{\frac{1}{4}} e^{\hat q-\hat p}\\
\left[\hat p,\hat q\right]=-h
\end{split}
\ee
Note that this choice actually embeds our $SL_q(2)$ into $SL_q(2)\otimes SL_{\tilde q}(2)$. Indeed, the generators
\be
\tilde K= e^{2\pi i H}, \quad \tilde E=E^{\frac{2\pi i }{h}},\quad \tilde F=F^{\frac{2\pi i}{h}}
\ee
commute with $K$, $E$ and $F$ and form a representation of $SL_q(2)$ with $\tilde q= e^{-\frac{4\pi^2}{h}}$.

In this case, it is simple to present the corresponding bases:
\be
e_{m,n,k}=\frac{H^m E^n \tilde E^k}{m!\prod\lm_{i=1}^n(1-q^i)\prod\lm_{j=1}^k(1-\tilde q^j)},\quad e^{m,n,k}=\bar H^m F^n\tilde F^k
\ee
Thus, the canonical element reads
\be
S=\sum\lm_{m,n,k\geq 0} e_{m,n,k}\otimes e^{m,n,k}=e^{H\otimes \bar H}\frac{(\tilde q^{-1}\tilde E\otimes \tilde F|\tilde q^{-1})_{\infty}}{(E\otimes F|q)_{\infty}}
\ee
Substituting explicit representation, one gets
\be
S_{1,2}=e^{ h^{-1}\hat q_1 \hat p_2}\;\Phi_h(\hat p_1+\hat q_2-\hat p_2)
\ee
where the quantum dilogarithm reads (in the domain $\mathop{\rm Re}\, h>0$):
\be\label{Hiko1}
\Phi_h(x)=\frac{(\tilde q^{-\frac{1}{2}}e^{\frac{2\pi i}{h}x}|\tilde q^{-1})_{\infty}}{(q^{\frac{1}{2}} e^{x}|q)_{\infty}}
\ee
Formulas (\ref{Hiko}) and (\ref{Hiko1}) give the ${\cal R}$-matrix that was used in \cite{Hiko} in constructing non-compact knot invariants, though it is different from Faddeev's ${\cal R}$-matrix (\ref{HikR}).

\section{RT formalism for infinite representations}

After we have constructed ${\cal R}$-matrices, we come to the second crucial ingredient of the RT formalism, to the notion of weighted trace, or, more generally, to turning operator. Remind that it is this ingredient that is so far unavailable in the case of infinite-dimensional Lie algebras. In this section we define the proper weighted trace in the case of non-compact group $SL_q(2)\otimes SL_{\tilde q}(2)$ and the ${\cal R}$-matrix (\ref{HF}) considered above, and explain how to make knot invariants using it.

\subsection{Weighted trace}

We are able to introduce a weighted trace fixing the second Reidemeister move:
\be\label{sec_Reid}
\qTr_z\; R_{s,s}\left[\begin{array}{cc}
    y_2 & z\\
    z & x_1 \\
\end{array}\right]\sim q^{\Omega_2(s)} \delta(x_1-y_2)
\ee

It is simpler to do in the \emph{momentum} space. Following \cite{KR,MS},
the weighted trace should read
\be
\qTr\; \star = \Tr\; q^{\pm 2H} \star
\ee
Since we actually work with the modular double, one needs to add the second copy:
\be
\qTr\; \star = \Tr\; q^{\pm 2H}{\tilde q}^{\pm 2\tilde H} \star
\ee
In the momentum space this trace reads
\be
\qTr\;\star =\int dk_1 dk_2\; \delta(k_1-k_2) e^{\pm 2\pi i Q k_1}\; \star
\ee
Thus, it is simple to verify eq.(\ref{sec_Reid}):
\be
\begin{split}
\int dz\;  e^{-2\pi i Q z} R_{s,s}\left[\begin{array}{cc}
    p_2 & z\\
    z & k_1 \\
\end{array}\right]=\delta(p_2-k_1)\int e^{-2\pi i Q z}dz\; \frac{e^{-i \pi  \left(k_1^2+z^2\right)}}{G_b\left(Q+i \left(z-k_1\right)\right)}\frac{ w_b\left(k_1+s\right) w_b(s-z)}{w_b\left(s-k_1\right) w_b(s+z) }=\\
=\delta(p_2-k_1) \frac{e^{i \pi  \left(2 k_1 s-k_1^2+s^2\right)} \Phi _b\left(s-k_1\right)}{\Phi _b\left(k_1+s\right)}\boxed{\int  e^{-2\pi i Q z}dz\; \frac{ \Phi _b\left(k_1+s+z\right)}{\Phi _b\left(-\frac{i Q}{2}+2 s+z\right) \Phi _b\left(z-\frac{i Q}{2}\right) }}
\end{split}
\ee
The boxed expression is known to be a $q$-counterpart of the reduced hypergeometric function integral at a fixed point (see \cite{AK}), hence
\be
\int  e^{-2\pi i Q z}dz\; \frac{ \Phi _b\left(k_1+s+z\right)}{\Phi _b\left(-\frac{i Q}{2}+2 s+z\right) \Phi _b\left(z-\frac{i Q}{2}\right) }\sim \Phi _b\left(k_1+s\right) \Phi _b\left(k_1- s\right)
\ee
Thus, one finally gets
\be
\boxed{\int dz\;  e^{-2\pi i Q z} R_{s,s}\left[\begin{array}{cc}
        p_2 & z\\
        z & k_1 \\
    \end{array}\right]=e^{2 i \pi  s^2} \delta(p_2-k_1)}
\ee

The Clebsh-Gordan coefficients ($V_{\alpha_1}\otimes V_{\alpha_2} \rightarrow \int d\alpha\; V_{\alpha_3}$) in the momentum space read
\be
\left[\begin{array}{c|cc}
    \alpha_3 & \alpha_2 & \alpha_1\\
    k_3 & k_2 & k_1
\end{array}\right]=\delta(k_3-k_2-k_1)\; C_{\alpha_3} \left[\begin{array}{cc}
 \alpha_2 & \alpha_1\\
 k_2 & k_1
\end{array}\right]
\ee
Thus, one can construct a "hat" operator $V_{\alpha_1}\otimes V_{\alpha_2} \rightarrow \IC$ as
\be
M_s(k_2,k_1):=\left[\begin{array}{c|cc}
    0 & s & s\\
    0 & k_2 & k_1
\end{array}\right]=\delta(k_2+k_1)\; C_{0} \left[\begin{array}{cc}
s & s\\
-k_1 & k_1
\end{array}\right]= \; \delta(k_2+k_1) e^{\pi i Q k_1}
\ee
and the trace is constructed with the two "hats"
\be
\qTr\; \star=\int dp\; dk\; dk'\; \left[\begin{array}{c|cc}
0 & s & s\\
0 & p & k
\end{array}\right]\left[\begin{array}{cc|c}
s & s & 0\\
 p & k' & 0
\end{array}\right]\star(k,k')=\int dk\; dk'\; \delta(k-k') e^{\pm 2\pi i Q k}\; \star
\ee

The Hopf algebra structure is given by the comultiplication
\be
\begin{split}
{\bf \Delta}(E)=E\otimes K+K^{-1}\otimes E\\
{\bf \Delta}(F)=F\otimes K+K^{-1}\otimes F\\
{\bf \Delta}(K)=K\otimes K
\end{split}
\ee
Hence, there is a simpler way to define the "hats", that is, to use the Hopf algebra structure:
\be
\begin{split}
{\bf \Delta}_{k_2,k_1}(E/F/H)M(k_2,k_1)=0\\
\int dk_1 dk_2\; \bar{M}(k_2,k_1){\bf \Delta}_{k_2,k_1}(E/F/H)\psi(k_2,k_1)=0,\; \forall \psi
\end{split}
\ee
The solution reads:
\be
\boxed{M_{s_1,s_2}(k_1,k_2)=\delta(k_1+k_2)e^{-\pi Q k_1}\delta_{s_1+s_2}}
\ee

\subsection{Knot invariants}

Now everything is ready to construct the non-compact knot invariant. Let us do this for the figure eight knot in order to illustrate the procedure. The knot is drawn at fig.3.

\begin{figure}[htbp]
    \begin{center}
        \includegraphics[scale=0.5]{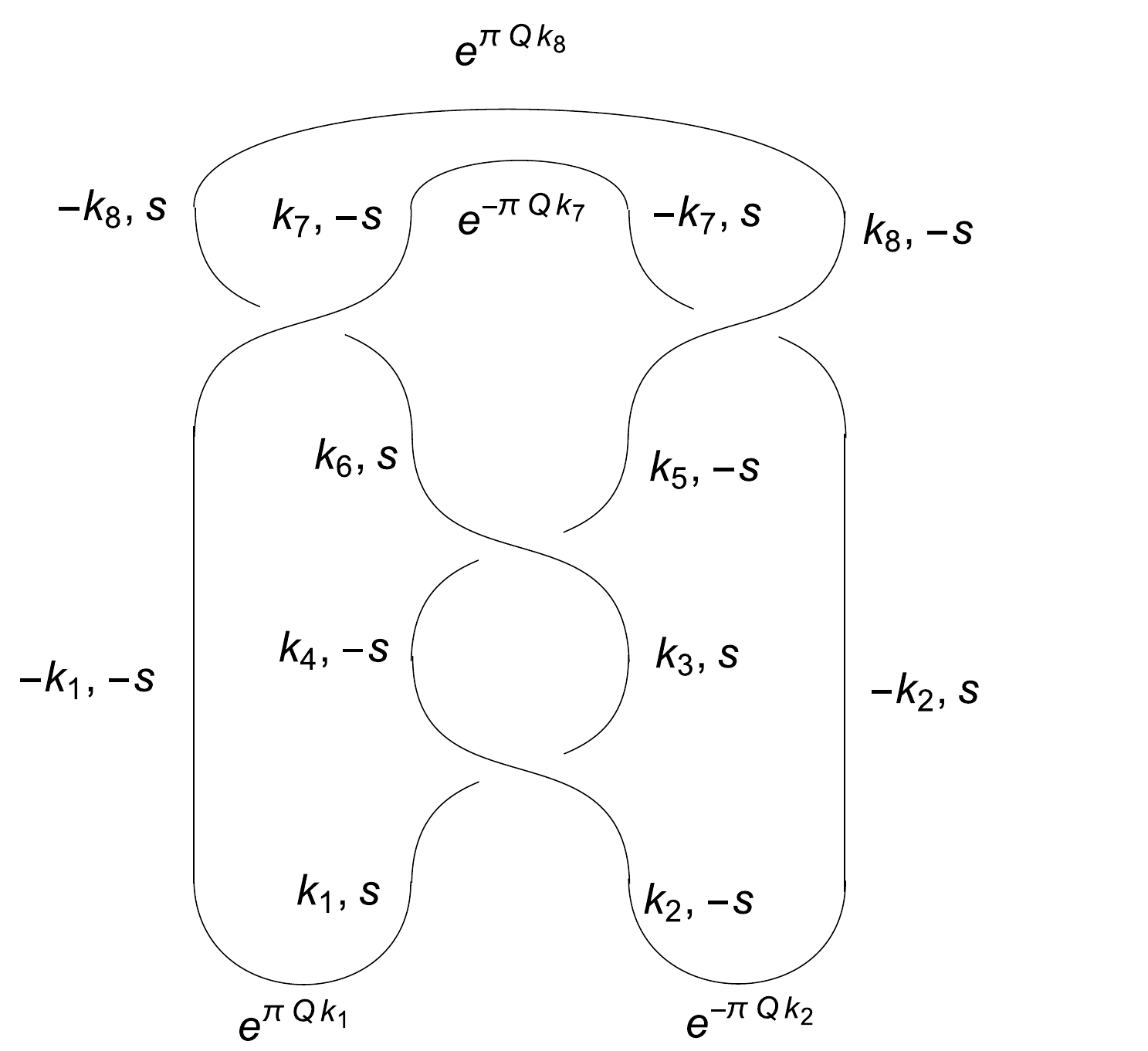}
        \caption{$4_1$ plane diagram \label{fig3}}
    \end{center}
\end{figure}

In accordance with the figure, one constructs the following expression:
\be
\langle 4_1\rangle=\int dk_1 dk_2 dk_3 dk_4 dk_5 dk_6 dk_7 dk_8\; e^{\pi Q (k_1-k_2-k_7+k_8)} R_{-s,s}^{-1}\left[\begin{array}{cc}
    k_4 & k_3\\
    k_2 & k_1 \\
\end{array}\right]R_{s,-s}^{-1}\left[\begin{array}{cc}
k_6 & k_5\\
k_3 & k_4 \\
\end{array}\right]\times \\ \times R_{-s,s}\left[\begin{array}{cc}
k_7 & -k_8\\
-k_1 & k_6 \\
\end{array}\right]R_{-s,s}\left[\begin{array}{cc}
k_8 & -k_7\\
k_5 & -k_2 \\
\end{array}\right]
=\int dy \frac{\Phi_b(s-y)}{\Phi_b(y)}e^{2\pi i s(2y-s)}
\ee
Similarly one can make the knot invariant out of any $2d$ knot diagram.

The same procedure can be applied for the other non-compact ${\cal R}$-matrix ({\bf iii}), see (\ref{Hiko}), (\ref{Hiko1}). However, in this case the ${\cal R}$-matrix depends on doubled number of parameters, but do not contain any manifest dependence on the spin $s$. It is not surprising, since this ${\cal R}$-matrix acts on the regular representation, \cite{BT}. Instead, there is a monodromy condition which fixes the representation and that is realized as added yet another integration with a $\delta$-function, in other words, as an additional constraint on the variables imposed, see \cite{Hiko}.

\section{Equations (A-polynomials) for knot invariants}

Let us now discuss what is the differences and similarities between the compact and non-compact invariants. First of all, we discuss the equations they satisfy. Let us again discuss the two simplest examples of the trefoil and figure eight knots, and the equations are difference equations in the spin variable.

\subsection{Trefoil}

We start with the Jones polynomial (\ref{Jtr}). It satisfies the following difference equation in spin of representation $r=N-1$, \cite{Gar}:
\be\label{difeq}
\mathds{J}_{N+1}+q^{6N+4}{1-q^{2N}\over 1-q^{2N+2}}\mathds{J}_N=q^{2N}{q^{4N+2}-1\over q^{2N+2}-1}
\ee
This equation can be rewritten in terms of operators
\be
\hat L \mathds{J}_N(q)=\mathds{J}_{N+1}(q),\ \ \ \ \hat M \mathds{J}_N(q)=q^N\mathds{J}_N(q)
\ee
in the form
\be\label{Apoltr1}
\left[q^2\hat M^6(\hat M^2-1)+(q^2\hat M^2-1)\hat L\right]\mathds{J}_N(q)=q^{2N}{q^{4N+2}-1\over q^{2N+2}-1}
\ee
One can make out of this equation a homogeneous difference equation, but of the second order
\be
\left[-{1\over q^2\hat M^2}{1-q^4\hat M^2\over 1-q^4\hat M^4}\hat L^2-\left(
{1\over \hat M^2}{1-q^2\hat M^2\over 1-q^2\hat M^4}-q^4\hat M^4{1-q^2\hat M^2\over 1-q^6\hat M^2}\right)\hat L
-q^4\hat M^4{1-\hat M^2\over 1-q^4\hat M^2}\right]\mathds{J}_N(q)=0
\ee
The operator in the l.h.s. of this equation is called quantum (or non-commutative) A-polynomial, since in the "classical" limit $q\to 1$ it coincides with the standard A-polynomial \cite{Apol} (a particular case of the AJ-conjecture \cite{Gar3}).

Equation (\ref{Apoltr1}) can be rewritten as an equation for the {\it un}reduced Jones polynomial $\mathds{K}_N(q)=[N]\mathds{J}_N(q)$:
\be\label{Apoltr}
\left(\hat L+q^{3}\hat M^6\right)\mathds{K}_N(q)={1-q^{4N+2}\over 1-q^{2N}}q^{3N-1}
\ee
The homogeneous equation of this equation has a simple solution, and it coincides with (\ref{Htr}) upon identification $N=x/\hbar$:
\be
\left(\hat L+q^{3}\hat M^6\right)\mathds{H}_N(q)=0,\ \ \Longrightarrow\ \ \mathds{H}_N(q)=e^{3\hbar N^2+i\pi N}\longrightarrow \mathds{H}(x)=e^{{3x^2\over\hbar} +i\pi {x\over\hbar}}
\ee 
Hence, the compact and non-compact invariants are different solutions of the same second order difference equation \cite{Z,Dim}. Or, to put it differently, they solve inhomogeneous first order difference equation and its homogeneous part respectively. In the meanwhile, their explicit forms look completely different: (\ref{Jtr}) and (\ref{Htr}). Still, one could try to look at the leading behaviour of the non-compact invariant as an asymptotics of the compact one at large spins \cite{Hik,3dAGT}.

\subsection{Figure eight}

The situation is completely the same for the figure eight knot. In this case, the Jones polynomial (\ref{Jfe}) satisfies the second order inhomogeneous difference equation (or a corresponding third order homogeneous equation)
\be
\left[q^4\hat M^4(1-\hat M^2)(1-q^6\hat M^4)-(q^2\hat M^2+1)(1-q^2\hat M^2-q^2\hat M^4-q^6\hat M^4-
q^6\hat M^6+q^8\hat M^8)(1-q^2\hat M^2)^2\hat L+\right.\\\left.
+q^4\hat M^4(1-q^2\hat M^4)(1-q^4\hat M^2)\hat L^2\right]\mathds{J}_N(q)=q^{2N+2}(1-q^{4N+6})(1-q^{4N+2})(1+q^{2N+2})
\ee
which homogeneous part acting on the {\it un}reduced polynomial is
\be
\left[q^{3}\hat M^4(1-q^6\hat M^4)-(1-q^4\hat M^4)(1-q^2\hat M^2-q^2\hat M^4-q^6\hat M^4-q^6\hat M^6+
q^8\hat M^8)\hat L+
q^{5}\hat M^4(1-q^2\hat M^4)\hat L^2\right]\mathds{K}_N(q)
=\hbox{...}
\label{Apolfe}
\ee
One could expect that the non-compact invariant (\ref{Hfe}) would again satisfy the corresponding homogeneous equation. This is, indeed, the case, though in order to check it, one needs some work.

\subsubsection{Ward identities}

To this end, we start with the non-compact integral invariant (\ref{Hfe}):
\be\label{HfeD}
\mathds{H}(x)=e^{-2\pi ix^2}\int dy\; e^{4\pi ixy}\frac{\Phi_b(x-y)}{\Phi_b(y)}
\ee
We need the following identities
\be\label{dl}
\Phi_b(z)=(1+qe^{2\pi bz})\Phi_b(z+ib)\\
\Phi_b(z)=(1+q^{-1}e^{2\pi bz})^{-1}\Phi_b(z-ib)
\ee
Let us also introduce the notation for ``expectation values'':
\be
A_{\pm}(U)= e^{-2\pi ix^2}\int dy\; e^{4\pi ixy}e^{\pm 2\pi by}\frac{\Phi_b(x-y)}{\Phi_b(y)}
\ee
We also define the operators:
\be
\hat L := e^{ib \frac{d}{dx}},\quad \hat M=e^{\pi bx}
\ee
First of all, one can easily get the Ward identity associated to the shift of integration variable $y\rightarrow y+ib$ in (\ref{HfeD}).
After applying identities (\ref{dl}) for dilogaritms, one gets the equation
\be
\Big(1+\hat M^{2} -\hat M^{4}\Big)\mathds{H}+q^{-1}\hat M^2A_-+qA_+=0
\ee
Similarly, one can get equations for $A_+(x)$ and $A_-(x)$ making shifts accordingly $x\rightarrow x+ ib$, $y\rightarrow y+ ib$ and $x\rightarrow x- ib$, the result reads:
\be
\hat L A_+=q\mathds{H}+ A_-
\\
\hat L^{-1} A_-=q^2\hat M^{-4}A_+ +qM^{-2}\mathds{H}
\ee
This system of three equations is easily reduced to the one equation for $\mathds{H}(x)$. After the rescaling
$\hat L\rightarrow q^{-1}\hat L$, which corresponds to transition from variable $N=r+1$ to variable $r$,
one gets the A-polynomial annihilating the non-compact invariant which coincides with (\ref{Apolfe}):
\be
\left[q^{3}\hat M^4(1-q^6\hat M^4)-(1-q^4\hat M^4)(1-q^2\hat M^2-q^2\hat M^4-q^6\hat M^4-q^6\hat M^6+
q^8\hat M^8)\hat L+
q^{5}\hat M^4(1-q^2\hat M^4)\hat L^2\right]\mathds{H}
=0\nn
\ee

\subsubsection{Compact vs non-compact invariants}

Note that in contrast with the trefoil case, the compact and non-compact invariants look more close in form. Indeed, one can convert the integral (\ref{HfeD}) into a sum over countable points. To this end, one has first to use the property
\be
\Phi_b(z)\Phi_b(-z)=e^{i\pi z^2}
\ee
and note that
\be
\Phi_b(z)={\prod\lm_{k=1}^{\infty} (1+q^{2k}e^{2\pi bz})\over \prod\lm_{k=-1}^{\infty} (1+\tilde q^{2k}e^{2\pi b^{-1}z})}=
\prod\lm_{k=1}^{\infty} (1+q^{2k}e^{2\pi bz}) f_{ib}(z)
\ee
where $f_T(z)$ denotes a $T$-periodic function in $z$. After changing the variables $x\rightarrow ib n$, and $y\rightarrow ib s$, the integral (\ref{HfeD}) reduces to
\be\label{App}
\mathds{H}(n)=q^{2n^2}\int ds\; q^{-4n s}\frac{(q^{2(n-s+1)}|q^2)_{\infty}}{(q^{2(s+1)}|q^2)_{\infty}}f_{ib}(s)\sim  q^{2n^2} \sum\lm_s  q^{-4n s}\frac{(q^{2(n-s+1)}|q^2)_{\infty}}{(q^{2(s+1)}|q^2)_{\infty}}
\ee
where the last transition is described in the Appendix and is done up to a periodic function. At the same time,
the Jones polynomial can be rewritten as
\be
\mathds{K}(n)\sim \sum\lm_s  q^{-2n s}\frac{(q^{2(n-s)}|q^2)_{\infty}}{(q^{2(n+s+1)}|q^2)_{\infty}}
\ee
These two expressions look quite similar, but not the same, though again one could try to identify leading behaviour of the first expression with an asymptotics of the second one \cite{Hik,3dAGT}. On the other hand, since these invariants satisfy the similar equation with different inhomogeneous parts, one may look at these expressions as just at different solutions to the same Ward identity, which usually encodes the basic information about the system.

Thus, one may ask to what extent the compact and non-compact invariants are independent, i.e. to what extent they are {\it different} invariants. The answer to this question seems to be still missing.

\section*{Acknowledgements}

This work was performed at the
Institute for Information Transmission Problems with the financial support of the Russian Science
Foundation (Grant No.14-50-00150).

\section{Appendix}

In this Appendix we derive claim (\ref{App}).

Consider a generic function labeled by a generic periodic measure $\mu$
\be
\phi_{\mu}\left[\begin{array}{c}
    \alpha_1,\ldots,\alpha_n\\
    \beta_1,\ldots,\beta_m\\
\end{array}\right](\gamma,\lambda|z)=\int d\sigma \; \mu(\sigma)\; z^{\sigma} q^{\gamma \sigma^2+\lambda \sigma}\frac{\prod\lm_{i=1}^n (\alpha_i q^{\sigma}|q)_{\infty}}{\prod\lm_{j=1}^m (\beta_j q^{\sigma}|q)_{\infty}}
\ee
This function satisfies a set of difference equations:
\be
\phi_{\mu}\left[\begin{array}{c}
    \alpha_1,\ldots,\alpha_n\\
    \beta_1,\ldots,\beta_m\\
\end{array}\right](\gamma,\lambda|z)=z\; q^{\gamma+\lambda}\;T_z^{2\gamma}\;\frac{\prod\lm_{j=1}^m(1-\beta_j T_z)}{\prod\lm_{i=1}^n(1-\alpha_i T_z)}\phi_{\mu}\left[\begin{array}{c}
\alpha_1,\ldots,\alpha_n\\
\beta_1,\ldots,\beta_m\\
\end{array}\right](\gamma,\lambda|z)\\
T_{\alpha_i}\phi_{\mu}\left[\begin{array}{c}
    \alpha_1,\ldots,\alpha_n\\
    \beta_1,\ldots,\beta_m\\
\end{array}\right](\gamma,\lambda|z)=(1-\alpha_i T_z)^{-1}\phi_{\mu}\left[\begin{array}{c}
\alpha_1,\ldots,\alpha_n\\
\beta_1,\ldots,\beta_m\\
\end{array}\right](\gamma,\lambda|z)\\
T_{\beta_j}\phi_{\mu}\left[\begin{array}{c}
    \alpha_1,\ldots,\alpha_n\\
    \beta_1,\ldots,\beta_m\\
\end{array}\right](\gamma,\lambda|z)=(1-\beta_j T_z)\phi_{\mu}\left[\begin{array}{c}
\alpha_1,\ldots,\alpha_n\\
\beta_1,\ldots,\beta_m\\
\end{array}\right](\gamma,\lambda|z)
\ee
where $T_x$ is the scaling operator: $T_x F(x) =f(q x)$.
These equations are also satisfied by the series
\be
F\left[\begin{array}{c}
    \alpha_1,\ldots,\alpha_n\\
    \beta_1,\ldots,\beta_m\\
\end{array}\right](\gamma,\lambda|z)=\sum\lm_s z^{s} q^{\gamma s^2+\lambda s}\frac{\prod\lm_{i=1}^n (\alpha_i q^{s}|q)_{\infty}}{\prod\lm_{j=1}^m (\beta_j q^{s}|q)_{\infty}}
\ee
Thus, one concludes that
\be
\phi_{\mu}\left[\begin{array}{c}
    \alpha_1,\ldots,\alpha_n\\
    \beta_1,\ldots,\beta_m\\
\end{array}\right](\gamma,\lambda|z)=f_{\mu}\left[\begin{array}{c}
\alpha_1,\ldots,\alpha_n\\
\beta_1,\ldots,\beta_m\\
\end{array}\right](z)\; F\left[\begin{array}{c}
\alpha_1,\ldots,\alpha_n\\
\beta_1,\ldots,\beta_m\\
\end{array}\right](\gamma,\lambda|z)
\ee
where the function $f$ is $\mu$-dependent, though it is $q$-periodic with respect to all the arguments.

\end{document}